\documentclass[preprint,a4paper,nofootinbib,showkeys,noindent,superscriptaddress]{revtex4-1}

\usepackage{amsmath,amssymb}
\usepackage{enumerate}
\usepackage{amsfonts}

\usepackage{subfigure}
\usepackage[hyperindex,colorlinks]{hyperref}

\usepackage{allpurpose}

\usepackage{tikz}
\usepackage{multirow}

\usepackage{array}
\newcolumntype{P}[1]{>{\centering\arraybackslash}p{#1}} 

\usepackage{slashed}

\usepackage{color}
\hypersetup{hidelinks,
        }

\begin{document}

\title{Generalised Amit-Roginsky model from perturbations of 3d quantum gravity}
\date{\today}

\author{Victor Nador}
\email[Email: ]{victor.nador@u-bordeaux.fr}
\affiliation{LaBRI, Univ. Bordeaux, 351 cours de la Lib\'eration, 33405, Talence, France}

\author{Daniele Oriti}
\email[Email: ]{daniele.oriti@physik.lmu.de}
\affiliation{Arnold Sommerfeld Center for Theoretical Physics, Ludwig-Maximilians-Universit\"at, Theresienstra\ss e 37, 80333, M\"uchen, Germany}

\author{Xiankai Pang}
\email[Email: ]{Xiankai.Pang@physik.uni-muenchen.de}
\affiliation{Arnold Sommerfeld Center for Theoretical Physics, Ludwig-Maximilians-Universit\"at, Theresienstra\ss e 37, 80333, M\"uchen, Germany}

\author{Adrian Tanasa}
\email[Email: ]{ntanasa@u-bordeaux.fr}
\affiliation{LaBRI, Univ. Bordeaux, 351 cours de la Lib\'eration, 33405, Talence, France}
\affiliation{LIPN, Univ. Sorbonne Paris Nord, Villetaneuse, France}
\affiliation{H. Hulubei Nat. Inst. Phys. Nucl. Engineering, P.O.Box MG-6, 077125, Magurele, Romania}

\author{Yi-Li Wang}
\email[Email: ]{Wang.Yili@physik.uni-muenchen.de}
\affiliation{Arnold Sommerfeld Center for Theoretical Physics, Ludwig-Maximilians-Universit\"at, Theresienstra\ss e 37, 80333, M\"uchen, Germany}

\begin{abstract}
  A generalised Amit-Roginsky vector model in flat space is obtained as the effective dynamics of pertubations around a classical solution of the Boulatov group field theory for 3d euclidean quantum gravity, extended to include additional matter degrees of freedom. By further restricting the type of perturbations, the original Amit-Roginsky model can be obtained. This result suggests a general link (and possibly a unified framework) between two types of tensorial quantum field theories: quantum geometric group field theories and tensorial models for random geometry, on one hand, and melonic-dominated vector and tensorial models in flat space, such as the Amit-Roginsky model (and the SYK model), on the other hand.
\end{abstract}


\maketitle
\tableofcontents

\section{Introduction}

Random matrix models ~\cite{David:1985nj,Ambjorn:1985az,DiFrancesco:2004qj, DiFrancesco:1993cyw} are, in their simplest formulation, 
$0$-dimensional 
field theories of an $N\times N$ (Hermitian) matrix $M_{ij}$ successfully employed to define $2$-dimensional euclidean quantum gravity, based on the fact that their perturbative expansion generates a sum over random surfaces weighted by purely combinatorial amplitudes corresponding to a simplicial gravity path integral on the triangulation dual to each matrix Feynman diagram. They have been generalized to matrix field theories in flat space, by the addition of suitable flat space coordinates, and used to describe, for example, large-N regimes of non-abelian gauge theories. Both finite matrix models and matrix field theories have found innumerable applications in mathematical and theoretical physics. 

A different kind of  
generalization is to define tensorial models producing, in their perturbative expansion, a sum over higher-dimensional lattices. Tensorial models in $d$ dimensions are obtained by replacing the matrix field $M$ by a tensor field with $d$ indices $M_{ij}\to T_{{i_1}..{i_d}}$ 

Such tensorial generalization was proposed already 30 years ago in a random geometry context ~\cite{Ambjorn:1990ge,Sasakura:1990fs,Godfrey:1990dt}, and soon adapted to the quantum geometric one for the description of topological quantum field theories \cite{Boulatov:1992vp,Ooguri:1992eb}, with 3d quantum gravity being a special case. The same quantum geometric models, under the label of group field theories, became central to formulate 4d quantum gravity in the context of spin foam models and canonical loop quantum gravity \cite{DP-F-K-R,P-R, P-R2}.
In this quantum gravity context, both as purely combinatorial random geometric models, and as richer quantum geometric ones, they represent nowadays a very promising and quickly developing area of research \cite{Freidel:06, Oriti:2011jm, Krajewski:2011zzu, Carrozza:2016vsq, Oriti:2016acw, Gielen:2016dss}.
The simplest example of such quantum geometric tensorial field theories is the so-called Boulatov model ~\cite{Boulatov:1992vp}, where
the r\^ole of the matrix indices is played here by group elements $g_1,g_2,g_3 \in SU(2)$. In these theories, the tensors $T_{i_1...i_d}$ of simple tensor models are replaced by fields $T(g_1,...,g_d)$ on a Lie group manifold $G^d$, having the local symmetries of gravitational theories in mind. 

More recently, tensorial field theories have proven to define very rich and interesting quantum field theories in flat space, again via the addition of suitable embedding coordinates; in particular, they define new conformal field theories, with many potential applications, e.g. to the AdS/CFT context \cite{Benedetti:2020yvb, Benedetti:2017fmp}.

The key mathematical fact that spurred much development in these models was the availability of analytic tools that allowed control over their perturbative expansion, despite the combinatorial intricacies.
Tensorial models, just like matrix models, admit a large $N$ expansion \cite{Gurau:2010ba, Gurau:2011xq, Dartois:2013he, Carrozza:2015adg, Tanasa:2015uhr, Tanasa:2012pm, Gurau:2019qag}.
The leading order in the tensor large $N$ expansion is given by a particular family of Feynman graphs called the \emph{melonic graphs}, which correspond to (special triangulations of) spherical topology. 
This analytic control has made possible the wealth of results on the renormalization group flow, both perturbative and non-perturbative, of tensorial field theories and group field theories \cite{Carrozza:2016vsq, Carrozza2017}, as well as the statistical analysis of critical behaviour \cite{Bonzom:2011, Baratin:2013rja}.

It is worth emphasizing here that the 
Sachdev-Ye-Kitaev (SYK) model \cite{Maldacena:2016hyu,Rosenhaus:2018dtp}
also enjoys the same melonic dominance in the large $N$ limit \cite{Bonzom:2018jfo}, with $N$ the number of fermionic fields of the SYK model.


The Amit-Roginsky (AR) model~\cite{Amit:1979ev} (see also \cite{Benedetti:2020iku}) describes a vector field theory 
whose coupling constant is proportional to an $SU(2)$ $3j$-symbol. 
This model also has a large $N$ expansion and one can prove that it exhibits a melonic limit, just like tensor models, where $N=2j+1$ is the dimension of the irreducible vector representation, and can thus be understood as a special (and particularly simple) element of tensorial vector field theories. 


Together, tensorial models of random and quantum geometry, and tensorial field theories in flat space, can be seen as part of a broader framework of {\it tensorial group field theories} (TGFTs), sharing key mathematical features and techniques, while remaining flexible enough to allow for a large variety of possible physical applications. However, the two classes of models have remained quite separate, so far.
The present work establishes the first explicit link between them.

A crucial ingredient will be, from the quantum geometric side of the story, the addition of matter degrees of freedom to the quantum geometric ones, also inspired by recent work on the extraction of a relational cosmological dynamics from group field theory \cite{Oriti:2016qtz, Marchetti:2020qsq, Marchetti:2021gcv}.
As a candidate of quantum gravity, the inclusion of matter is of course crucial for TGFTs. Work in this direction has followed two main routes, rather disconnected. On the one hand, non-commutative scalar field theories have been extracted, by interpreting the Lie group domain of quantum geometric models as a curved momentum space, as perturbations over classical solutions of group field theories \cite{Fairbairn:2007sv, Girelli:2009yz}, producing an \lq emergent matter\rq description from the same quantum degrees of freedom having a pre-geometric interpretation. On the other hand, matter degrees of freedom (or, maybe more properly, \lq pre-matter\rq degrees of freedom) have been added to the quantum geometric ones, so to produce a lattice path integrals for the coupling of gravity and matter at the level of the Feynman amplitudes of GFT models \cite{Li:2017uao, Oriti:2006jk, Fairbairn:2006dn}. These additional degrees of freedom are also instrumental for the definition of relational observables with a local spacetime interpretation, in group field theory cosmology, as mentioned. 
Both strategies turn out to be relevant for linking the quantum geometric Boulatov model to the Amit-Roginsky model, in this work, with the latter arising as the effective dynamics of quantum geometric perturbations, but with the additional degrees of freedom interpret as matter frames in the quantum geometric setting playing the role of flat space coordinates in the resulting Amit-Rogisnky model.


\medskip

In this paper, the classical solutions of the equation of motion of the Boulatov model regularised via a heat-kernel approach are investigated. We exhibit an explicit solution of these equations with the $3j$ symbol of $SU(2)$ and study $2$-dimensional perturbations around this solution. We then give explicit conditions on these perturbations to give rise to an AR-like effective action - with additional summation over spin indices with respect to the original AR model. This shows that the AR model can be seen as a 
perturbation around classical solutions of the Boulatov model, thus giving the anticipated explicit link between two types of tensorial models.\footnote{We notice here that the possibility of deriving the AR vector model from perturbations over special solutions of tensorial models was suggested also in \cite{Benedetti:2019sop}. A $SO(3)$-invariant classical solution in the form of a 3j-symbol was identified for a $O(N)^3$-symmetric tensor model in the large-N limit, corresponding to an interesting pattern of symmetry breaking. Although the model considered in that paper is different from the extended Boulatov TGFT model we analyse, and no direct quantum gravity interpretation is immediately available, the general mechanism is quite similar to the one we study. A more in-depth comparative analysis of the two settings would be very interesting.} 

The paper is organised as follows. A brief review of the Boulatov model is given in section~\ref{sec:bolatovreview}. In the following section we recall the definition of the AR model; then, we study the condition on the perturbations of classical solutions of the equations of motion of the Boulatov model necessary to recover an AR-like action as an effective action, which is then explicitly derived. Section~\ref{melonicdominance} discusses the existence of a melonic dominance for our effective action. While it is unsettled whether melonic dominance is preserved in the most general setting, we exhibit additional conditions that ensure this property. Finally, we offer some conclusions and perspectives.

\section{Boulatov model} \label{sec:bolatovreview}
\subsection{A short review on the Boulatov GFT model}

Quantum geometric TGFTs, or GFTs, \cite{Freidel:2005qe,Oriti:2006se,Krajewski:2011zzu,Oriti:2011jm,Oriti:2013aqa} are field theories whose dynamical field depends on $n$ points $g_i$ of a Lie group $G$. 
The group elements $g_i$ can be interpreted as discrete parallel transports of a gravitational connection, {\it i.e.} of a $G$-vector bundle. 
The Boulatov model~\cite{Boulatov:1992vp} is a $3$D GFT model with field $T(g_1,g_2,g_3) :G^{3}\to \mathbb{C}$, where $G=SU(2)$. 
The field is invariant under 
\begin{equation}
    T(g_1h,g_2h,g_3h)=T(g_1,g_2,g_3)\hspace{10pt}\forall h\in SU(2). 
    \label{right_inv}
\end{equation}
and satisfies the reality condition \cite{Fairbairn:2007sv}
\begin{equation}
    T(g_1,g_2,g_3)=\bar{T}(g_3,g_2,g_1). \label{reality}
\end{equation}
The original Boulatov model~\cite{Boulatov:1992vp} further requires cyclic symmetry in the group elements $g_i$. But this property plays no role in this paper, so we do not discuss it further.

The action of the Boulatov model is non-local and it writes~\cite{Boulatov:1992vp}
\begin{align}
          S[T]&=\frac{\mu^2}{2}\int \dd g_1 \dd g_2 \dd g_3T(g_1,g_2,g_3)\bar{T}(g_1,g_2,g_3) \nonumber \\
      &-\frac{\lambda}{4!}\int \prod_{i=1}^6 \dd g_i T(g_1,g_2,g_3) T(g_3,g_5,g_4)T(g_4,g_2,g_6)T(g_6,g_5,g_1), \label{eq:actionTgroup}
\end{align}
where $\mu$ is the \lq mass\rq of the field (simply the coupling of the quadratic non-derivative term) and $\lambda$ is the coupling constant of the quartic interaction. The connection to simplicial geometries is elucidated by a suitable graphical interpretation of the elements in the action.
The field $T(g_1,g_2,g_3)$ represents a triangle, with three group elements associated with its three edges, and the interaction contains four triangles glued along shared edges (thus sharing the same group element) forming a tetrahedron, which is the building block of a $3D$ simplicial lattice, likes those generated as dual to the Feynman diagrams of the model in its perturbative expansion.

The equation of motion of the field $T(g_1,g_2,g_3)$ reads
\begin{equation}
      \mu^2T(g_3,g_2,g_1)=\frac{\lambda}{3!}\int\dd g_4\dd g_5\dd g_6T(g_3,g_5,g_4)T(g_4,g_2,g_6)T(g_6,g_5,g_1). \label{eq:boulatoveqTgroup}
\end{equation}
This provides a description of the GFT model in a \emph{group representation}. By generalised Fourier transforms, GFTs can also be written in terms of a \emph{spin representation}.

As a function of $SU(2)^{\otimes3}$, the field $T$ can be expanded in terms of Wigner matrices $D^{j_i}_{m_in_i}(g_i)$ via the Peter-Weyl theorem \cite{Makinen:2019rou,Martin-Dussaud:2019ypf}. Considering the invariance~\eqref{right_inv}, this decomposition takes the form 
\begin{equation}
      T(g_1,g_2,g_3)=\sum_{\{j,m,n\}}T^{m_1m_2m_3}_{j_1j_2j_3}\prod_{i=1}^{3}\sqrt{2j_i+1}D^{j_i}_{m_in_i}(g_i)\mat{ccc}{j_1&j_2&j_3\\n_1&n_2&n_3}, \label{eq:Tdecompos}
\end{equation}
with $\displaystyle \mat{ccc}{j_1&j_2&j_3\\n_1&n_2&n_3}$ the Wigner's $3j$ symbol of $SU(2)$. 
The sum on ${\{j\}}$ denotes the summation over $j_1,~j_2$ and $j_3$ (resp. for $\{m\}$ and $\{n\}$).
The coefficients $T^{m_1m_2m_3}_{j_1j_2j_3}$ can be computed using the orthogonality of Wigner matrices as
\begin{equation}
      T^{m_1m_2m_3}_{j_1j_2j_3}=\int \dd g_1 \dd g_2 \dd g_3 \sum_{\{n\}}T(g_1,g_2,g_3)\prod_{i=1}^3\sqrt{2j_i+1}\bar{D}^{j_i}_{m_in_i}(g_i)\mat{ccc}{j_1&j_2&j_3\\n_1&n_2&n_3}, \label{eq:Tdecomposcoeff}
\end{equation}
Using this decomposition, the integral over the Wigner matrix can be performed explicitly and the Boulatov action~\eqref{eq:actionTgroup} in spin representation reads \cite{Boulatov:1992vp}
\begin{equation}
      S_B[T] = \sum_{j_1,j_2,j_3} \frac{\mu^2}{2}|T^{m_1,m_2,m_3}_{j_1,j_2,j_3}|^2 - \frac{\lambda}{4!} \sum_{j_1,..,j_6} \sixj{j_1}{j_2}{j_3}{j_4}{j_5}{j_6}  T^{4_{6j}}, \label{eq:actionTspin}
\end{equation}
where the kinetic term is
\begin{equation}
|T^{m_1,m_2,m_3}_{j_1,j_2,j_3}|^2=\sum_{\substack{j_1,j_2,j_3\\m_1,m_2,m_3}}(-1)^{\sum_{i=1}^3(j_i-m_i)}T^{m_1,m_2,m_3}_{j_1,j_2,j_3}T^{-m_1,-m_2,-m_3}_{j_1,j_2,j_3},
\end{equation}
and the term $T^{4_{6j}}$ encodes the contraction of the magnetic indices $m_i$ of the field paralleling the contraction pattern of 3j-symbols to give the $6j$ symbol, {\it i.e.}
\begin{equation}
      T^{4_{6j}}=\sum_{\{j,m\}}(-1)^{\sum_{i=1}^{6}(j_i-m_i)}T^{-m_1,-m_2,-m_3}_{j_1j_2j_3}T^{m_3,m_5,-m_4}_{j_3j_5j_4}T^{m_4,m_2,-m_6}_{j_4j_2j_6}T^{m_1,-m_5,m_1}_{j_6j_5j_1}. \label{eq:T6jdef}
\end{equation}
In this form, the equation of motion~\eqref{eq:boulatoveqTgroup} now becomes
\begin{equation}
      \mu^2T^{m_1,m_2,m_3}_{j_1,j_2,j_3} = \frac{\lambda}{3!} \sum_{j_4,j_5,j_6} \sixj{j_1}{j_2}{j_3}{j_4}{j_5}{j_6} T^{4_{6j}}_{\backslash \{m_1,m_2,m_3\}}, \label{eq:boulatoveqTspin}
\end{equation}
where
\begin{equation}
      T^{4_{6j}}_{\backslash \{m_1,m_2,m_3\}}=\sum_{m_4,m_5,m_6}(-1)^{\sum_{i=4}^6(j_i-m_i)}T^{m_3,m_5,-m_4}_{j_3j_5j_4}T^{m_4,m_2,-m_6}_{j_4j_2j_6}T^{m_6,-m_5,m_1}_{j_6j_5j_1}.
\end{equation}
is the field $T$ where the three magnetic indices $m_1$,$m_2$ and $m_3$ are not summed on.

In the rest of this article we will use this spin representation~\eqref{eq:actionTspin} of the Boulatov action.

Finally, before we discuss how matter degrees of freedom are included in the Boulatov model, let us recall some facts concerning its interpretation as a model for 3d euclidean quantum gravity. Its Feynman amplitudes are given by lattice gravity path integrals corresponding to a discretization of 1st order Palatini 3d gravity on the lattices dual to the Feynman diagrams. Equivalently, they correspond to the Ponzano-Regge spin foam amplitudes, known to correspond to a state sum formulation of the same quantum theory. In absence of matter, this quantum theory only describe flat 3d geometries and the partition function, for given lattice, counts the moduli space of flat connections for the given topology. The Boulatov model extends thus this quantum geometric content with a sum over lattices of all topologies (all possible gluings of 3-simplices), including pseudomanifold configurations. The quantum geometric effect of this additional sum is not fully understood. While the sum over lattices with the same topology is most likely irrelevant from the physical point of view, and, once controlled, should give at most a rescaling of the amplitudes, the sum over different topologies may have more interesting physical consequences. Tree level amplitudes, however, should not encode such topological effects, thus it is natural to interpret classical solutions of the Boulatov model as still corresponding to flat space. Clearly, further work is needed to improve our understanding of these issues. 

\subsection{Matter degrees of freedom}

GFTs are not usual QFTs describing a theory \emph{on} spacetime, but QFTs \emph{of} spacetime, tentatively describing its quantum building blocks and their dynamics \cite{Oriti:2011jm}. Their dynamical fields do not live, accordingly, on a manifold interpreted as spacetime, and on which the usual metric and matter fields of GR and standard model live. Such spacetime manifold simply does not appear in the fundamental formulation of the theory, as one does not find coordinates and directions on such manifold. 

According to the relational strategy for the construction of diffeomorphism-invariant observables in classical and quantum gravity~\cite{Rovelli:1990ph,Rovelli:2004tv}, spacetime localization should be defined in terms of appropriately chosen dynamical degrees of freedom, {\it internal} to the theory, rather than absolute external directions. For example, matter coupled to gravity can play the role of a physical reference frame~\cite{Rovelli:1990ph}, i.e. of rods and clocks. While different choices of matter can be used to fill that role, the simplest framework is to use free massless (minimally coupled) scalar fields $\chi_i$~\cite{Oriti:2016qtz,Oriti:2017}. 

In three dimensions, one needs three scalar fields, and they can be combined into a vector $\vec{\chi}=(\chi_1,\chi_2,\chi_3)$, to be added to the GFT data to localize in space and time, in a continuum approximation, GFT observables and their dynamics.

We now exhibit a specific construction extending the Boulatov model to include such matter degrees of freedom. Other constructions can be found in the cited GFT literature.
Requiring the theory to be invariant under translations $\chi_i\to\chi_i+a_i$ allows for a kinetic term in the action~\eqref{eq:actionTgroup} defined as $\displaystyle \nabla =\left(\frac{\partial }{\partial \chi_1},\frac{\partial }{\partial \chi_2},\frac{\partial }{\partial \chi_3}\right)$ and thus extends $T(g_1,g_2,g_3)$ to $T(g_1,g_2,g_3;\vec{\chi}):SU(2)^3\times\mathbb{R}^3\to\mathbb{C}$. The new action writes~\footnote{Note that this action should not be confused with that of a dynamical Boulatov model of~\cite{Geloun:2013} where a Laplace-Beltrami operator acts on the group manifold.}
\begin{align}
    S[T]&=\int [\dd g]^3\dd^3\vec{\chi}\left[\frac{1}{2}\nabla T(g_1,g_2,g_3;\vec{\chi})\nabla \bar{T}(g_1,g_2,g_3;\vec{\chi})+ \frac{\mu^2}{2}T(g_1,g_2,g_3;\vec{\chi})\bar{T}(g_1,g_2,g_3;\vec{\chi})\right] \nonumber \\
      &-\frac{\lambda}{4!}\int \prod_{i=1}^6 \dd g_i\dd^3\vec{\chi} T(g_1,g_2,g_3;\vec{\chi}) T(g_3,g_5,g_4;\vec{\chi})T(g_4,g_2,g_6;\vec{\chi})T(g_6,g_5,g_1;\vec{\chi}).\label{eq:actionTchigroup}
\end{align}
   This yields a modified equation of motion: 
\begin{align}
      &\nabla^2 T(g_3,g_2,g_1;\vec{\chi})+\mu^2T(g_3,g_2,g_1;\vec{\chi})\nonumber\\
  &=\frac{\lambda}{3!}\int\dd g_4\dd g_5\dd g_6T(g_3,g_5,g_4;\vec{\chi})T(g_4,g_2,g_6;\vec{\chi})T(g_6,g_5,g_1;\vec{\chi}). \label{eq:boulatoveqTchigroup} 
\end{align}
The corresponding action in spin representation writes
\begin{align}
        S_B[T(\vec{\chi})] &= \sum_{j_1,j_2,j_3}\int\dd^3\vec{\chi}\left[\frac{1}{2}\left|\nabla T^{m_1,m_2,m_3}_{j_1,j_2,j_3}(\vec{\chi})\right|^2+ \frac{\mu^2}{2}\left|T^{m_1,m_2,m_3}_{j_1,j_2,j_3}(\vec{\chi})\right|^2\right. \nonumber \\
    & \left.- \frac{\lambda}{4!} \sum_{j_1,..,j_6} \sixj{j_1}{j_2}{j_3}{j_4}{j_5}{j_6}  \int\dd^3\vec{\chi}T(\vec{\chi})^{4_{6j}}\right], \label{eq:actionTchispin}
\end{align}
leading to the following equation of motion:
\begin{align}
      \nabla^2 T^{m_1,m_2,m_3}_{j_1,j_2,j_3}(\vec{\chi})+\mu^2T^{m_1,m_2,m_3}_{j_1,j_2,j_3}(\vec{\chi}) = \frac{\lambda}{3!} \sum_{j_4,j_5,j_6} \sixj{j_1}{j_2}{j_3}{j_4}{j_5}{j_6} T(\vec{\chi})^{4_{6j}}_{\backslash \{m_1,m_2,m_3\}}. \label{eq:boulatoveqTchispin}
\end{align}

Before we take our next step in the derivation, we point out that TGFT models of the above \lq extended\rq type, including both local and non-local (tensorial) directions have also been analysed, recently, from the point of view fo their renormalization group flow \cite{Geloun:2023ray} and their critical behaviour (at mean field level) \cite{Marchetti:2020xvf, Marchetti:2022nrf}.

\subsection{Classical homogeneous solutions to the Boulatov model}

We first exhibit a homogeneous classical solution of the Boulatov model, independent of $\vec{\chi}$. In the homogeneous restriction, equation~\eqref{eq:boulatoveqTchigroup} reduces to equation~\eqref{eq:boulatoveqTgroup}. The dependence on the scalar matter degrees of freedom $\chi_i$ will only enter perturbatively around this solution. A one-parameter family of solutions parametrized by normalised class functions $f:SU(2)\to \mathbb{C}$ was proposed in~\cite{Fairbairn:2007sv}, with associated field $T_f$ is given by
\begin{equation}
     T_f(g_1,g_2,g_3)=\mu\sqrt{\frac{3!}{\lambda}}\int \dd h\delta(g_1h)f(g_2h)\delta(g_3h), \label{eq:Tfsolgroup}
\end{equation}
where $\delta(g)$ is the Dirac delta function over the group $SU(2)$ such that
\begin{equation}
      \int \dd h \delta(h)=1, ~\int \dd h\delta(h)f(h)=f(I),
\end{equation}
with $I$ is the identity of $SU(2)$ group. 

The function $f(g)$ is normalised, {\it i.e.}
\begin{equation}
  \int \dd h f(h)^2=1. \label{eq:fnorcon}
\end{equation}
We can also write this solution in spin representation, substituting the solution \eqref{eq:Tfsolgroup} into the general Peter-Weyl coefficients \eqref{eq:Tdecomposcoeff}, to obtain
\iea{
  (T_f)_{j_1,j_2,j_3}^{m_1,m_2,m_3} = \mu\sqrt{\frac{3!}{\lambda}} \sqrt{d_{j_1}d_{j_3}} \sum_{l_2} f^{j_2}_{m_2,l_2} \threej{j_1}{j_2}{j_3}{m_1}{l_2}{m_3}, \label{eq:Tfsolspin}
}
where $f^{j}_{mn}$ is the coefficients in the Peter-Weyl decomposition of $f(g)$
\begin{equation}
     f^{j}_{mn}=\sqrt{2j+1}\int \dd g f(g)\bar{D}^{j}_{mn}(g), \label{eq:fdecomposcoeff}
\end{equation}
and the corresponding normalisation condition becomes
\iea{
  \sum_{j,m,n} (-1)^{m-n}f^j_{mn}f^j_{-m,-n}=1. \label{eq:fnorspin}
}

Before we move on, let us give some remark on this class of solutions and its special form which is regularised by ``heat kernel''. First, the classical solution\eqref{eq:Tfsolgroup} is asymmetrical in the group elements $g_i$ since $g_2$ plays a preferential role through $f$. Restricting attention to this special asymmetric solutions is thus a form of spontaneous symmetry breaking of the model. Second, the presence of Dirac delta function in ~\eqref{eq:Tfsolgroup} leads to divergences. For example, the action~\eqref{eq:boulatoveqTchigroup} is divergent when evaluated on this solution due to the factor $\delta(I)$ appearing. This can also be seen from its Peter-Weyl expansion
\begin{equation}
     \delta(g)= \sum_{j,m} (2j+1) D^j_{mm}(g).
\end{equation}
Thus we need to regularize our solution. This can be achieved by different methods. For example, one strategy is to introduce a sharp cut-off parameter $J$ in the Peter-Weyl expansion of $T(g_1,g_2,g_3)$, thus making the action finite. Here, we will instead use a \emph{heat kernel} regularization to make all quantities well-defined, at the cost of only having an approximate solution to the equations of motion. To do so, we introduce a new real parameter $\varepsilon$. For any function $f$ of $SU(2)$ with coefficients $f^j_{mn}$ in its Peter-Weyl expansion, we define its heat kernel regularization as ($d_j=2j+1$)
\begin{equation}
    f_\epsilon(g) = \sum\limits_{j,m,n} \sqrt{d_j} f^j_{mn} D^j_{mn}(g) \ee^{-\varepsilon C_j}
\end{equation}
with $C_j$ is the Casimir of the spin $j$ representation of $SU(2)$. This function is well-defined for any $\epsilon>0$ and its leading order when $\varepsilon \rightarrow 0$ is the initial function $f$. In particular, for the Dirac delta function of $SU(2)$, its heat kernel regularization is
\begin{equation}
    \delta_\varepsilon(g)=\sum_{j,m}d_j D^j_{mm}(g)\ee^{-\varepsilon C_j}.
\end{equation}

Note that this function is not normalised. If we denote its norm as $\alpha_\epsilon^{-2}$, the normalised function associated to $\delta_\varepsilon$ is ($d_j=2j+1$)
\begin{equation}
    \Delta_\varepsilon(g) =  \alpha_\varepsilon \sum\limits_{j,m,n} \sqrt{d_j}(\Delta_\varepsilon)^j_{mn} D^j_{mn}(g) \ee^{-\varepsilon C_j},
\end{equation}
where the Peter-Weyl coefficients $(\Delta_\varepsilon)^j_{mn}$ has the form
\begin{equation}
     (\Delta_\varepsilon)^j_{mn}=\alpha_\varepsilon \sqrt{d_j} \delta_{mn} \ee^{-\varepsilon C_j}. \label{eq:Deltavarepsiloncoef}
\end{equation}

Using $\Delta_\varepsilon(g)$, we can build now a regularized and symmetric field
\begin{equation}
     T_\varepsilon(g_1,g_2,g_3)=\mu\sqrt{\frac{3!}{\lambda}}\int \dd h\delta_\varepsilon(g_1h)\Delta_\varepsilon(g_2h)\delta_\varepsilon(g_3h)=\mu\alpha_\varepsilon\sqrt{\frac{3!}{\lambda}}\int \dd h\delta_\varepsilon(g_1h)\delta_\varepsilon(g_2h)\delta_\varepsilon(g_3h).\label{eq:Tepsilonsolgroupreg}
\end{equation}

However, $T_\varepsilon(g_1,g_2,g_3)$ is only an approximate solution of the homogeneous equation of motion, i.e. it is a solution at leading order in $\varepsilon$. The coefficients of its Peter-Weyl expansion are given by 
\begin{equation}
    \left(T_\varepsilon\right)^{m_1m_2m_3}_{j_1j_2j_3}= \mu\alpha_{\varepsilon}\sqrt{\frac{3!}{\lambda}}\prod_{i=1}^3\sqrt{d_{j_i}}\ee^{-\varepsilon C_{j_i}} \mat{ccc}{j_1&j_2&j_3\\m_1&m_2&m_3}. \label{eq:PW_eps}
\end{equation}
In particular, when $\varepsilon \rightarrow 0$ the coefficients of $T$ are given by the $3j$ symbol, which is a (regularized) classical solution to the Boulatov model.\\

In the following calculation, we will use the solution (\ref{eq:Tfsolgroup}) and will briefly illustrate the special case (\ref{eq:Tepsilonsolgroupreg}) separately.

\section{Amit-Roginsky-like model from perturbations around classical Boulatov solutions}
\label{sec:emergenceAR}

In this section, we obtain an AR-like action from the Boulatov GFT action by considering specific perturbations around the classical solution constructed in the previous section.

The AR model~\cite{Amit:1979ev} is a cubic field theory of a vector field $\phi$ self-coupled through the $3j$ symbol for a fixed value of the spin $j$. Its action is
\begin{align}
  S_{AR}[\phi]=&\int\dd^dx\left\{\frac{1}{2}\sum_m(-1)^{j-m}\left[(\nabla \phi^j_m)(\nabla \phi^j_{-m})+\mu\phi^j_m\phi^j_{-m}\right]\right. \nonumber \\
  &+\left.\sum_{m_1,m_2,m_3}\frac{\lambda}{3!}\sqrt{2j+1}\mat{ccc}{j&j&j\\m_1&m_2&m_3}\phi^j_{-m_1}\phi^j_{-m_2}\phi^j_{-m_3}\right\}, \label{eq:actionAR}
\end{align}
where $\displaystyle \nabla$ is the gradient operator.

It was recently pointed out in~\cite{Benedetti:2020iku} that the large $N(=2j+1)$ limit of the AR model is given by the~\emph{melonic} graphs. As mentioned in the introduction, this feature is shared with $0$-dimensional tensor models~\cite{Dartois:2013he, Carrozza:2015adg} and topological GFTs as well. 

\subsection{Perturbations over homogeneous Boulatov solution}

Following~\cite{Fairbairn:2007sv}, we consider two-dimensional perturbations over the Boulatov model, which depend on matter reference frame $\vec{\chi}$. The field becomes
\begin{equation}
    T_\psi(g_1,g_2,g_3;\vec{\chi})=T_f(g_1,g_2,g_3)+\xi \psi(g_1,g_3;\vec{\chi}), \label{eq:2dpert_solution}
\end{equation}
where $T_f(g_1,g_2,g_3)$ is the solution to the equation of motion given by equation~\eqref{eq:Tfsolgroup} with \eqref{eq:Tepsilonsolgroupreg} a special case, and $\psi(g_1,g_3;\vec{\chi})$ is a $2$D-perturbation with 
$\xi$ a real parameter 
$0<\xi\ll 1$. The Peter-Weyl coefficients of the perturbation are given by
\begin{align}
      \psi^{m_1m_2m_3}_{j_1j_2j_3}(\vec{\chi})&=\sum_{\{n\}}\int[\dd g]^3\psi(g_1,g_3;\vec{\chi})\prod_{i=1}^3\sqrt{2j_i+1}\bar{D}^{j_i}_{m_in_i}\mat{ccc}{j_1&j_2&j_3\\n_1&n_2&n_3} \nonumber \\
  &\equiv \delta^{j_2,0}\delta_{m_2,0}\delta^{j_1,j_3}\sqrt{2j_1+1}\psi^{j_1}_{m_1,m_3}(\vec{\chi}).
\end{align}
 In order to obtain the equation above, we used the fact that $j_2=0$ (see equation~\eqref{eq:3jj2eq0} in the appendix~\ref{sec:su2recoupling}). The scaling factor $\sqrt{2j_1+1}$ is introduced for later convenience. The Peter-Weyl coefficients of the perturbed solution write
\begin{equation}
      (T_\psi)^{m_1m_2m_3}_{j_1j_2j_3}(\vec{\chi})=T^{m_1m_2m_3}_{j_1j_2j_3}+\xi \delta^{j_2,0}\delta_{m_2,0}\delta^{j_1,j_3}\psi^{j_1}_{m_1,m_3}(\vec{\chi}). \label{eq:Tpsispin}
\end{equation}

Substituting~\eqref{eq:Tpsispin} into the action~\eqref{eq:actionTchispin}, we get the action for the perturbed solution
\begin{equation}
      S_B[T_\psi(\vec{\chi})]=S_B[T] + \xi^2\cdot S_{\mathrm{eff}}[\psi]+\mathcal{O}(\xi^4),
\end{equation}
where the first order in $\xi$ vanishes since $T_f$ is a solution to the equation of motion. The action $S_{\mathrm{eff}}[\psi]$ represents the effective action of the perturbation field $\psi^j_{mn}$ and contains corrections up to $\xi$. Therefore, $\xi^2 S_{\mathrm{eff}}[\psi]$ contains corrections up to order $\xi^3$. 

In the following subsection, we develop each term arising from the Boulatov model in the effective action and give sufficient conditions on the coefficients $(T_\epsilon)^{m_1m_2m_3}_{j_1j_2j_3}$ such that the effective action $S_{\mathrm{eff}}[\psi]$ takes the form of an AR-like action. Since the AR model involves a vector field transforming in a representation of $SU(2)$ and thus carrying only one magnetic index $m$. Hence, we will specialize the perturbations to
\begin{equation}
      \psi^{j_1}_{m_1m_3}(\vec{\chi})=\sum_{m}\sqrt{2j_1+1}\phi^{j_1}_{m}(\vec{\chi})\mat{ccc}{j_1&j_1&j_1\\m_1&m&m_3}.
      \label{eq:psiphirelation}
\end{equation}
and check that this particular choice of perturbations satisfies all the required conditions.

\subsection{Conditions for the emergence of an Amit-Roginsky-like model} 

In order to simplify the notations, we omit from now on to explicitly write the dependency on the vector $\vec{\chi}$, which should always be assumed. 

\subsubsection{Quadratic terms}

Substituting perturbation~\eqref{eq:Tpsispin} into the Boulatov action~\eqref{eq:actionTchispin}, the quadratic term in $\xi$ receives three kinds of contributions.
The kinetic term of Boulatov model gives rise to one contribution of the form $\psi\psi$. Then, the interaction term gives two distinct type of contributions, depending on how the two perturbation fields are connected in the action. 

Schematically, these two terms can be represented as $TT\psi\psi$ when the two perturbation fields $\psi_{m_am_b}$ share one magnetic index, and $T\psi T\psi$ represents the terms that share none. They yield different contributions to the effective action.

\vspace{10pt}
\paragraph{Term $\psi\psi$\\}

The kinetic term $\sum_{j_1,j_2,j_3}\left|(T_\psi)^{m_1,m_2,m_3}_{j_1,j_2,j_3}(\vec{\chi})\right|^2$ of the Boulatov action
gives the following contribution to the effective action:
\begin{align}
      &\sum_{\substack{j_1,j_2,j_3\\m_1,m_2,m_3}}(-1)^{\sum_{i=1}^3(j_i-m_i)}\left[\delta^{j_2,0}\delta_{m_2,0}\delta^{j_1,j_3}\psi^{j_1}_{m_1,m_3}\right]\left[\delta^{j_2,0}\delta_{-m_2,0}\delta^{j_1,j_3}\psi^{j_1}_{-m_1,-m_3}\right]  \nonumber \\
  &=\sum_{\substack{j_1,m_1,m_3\\m,m'}}(-1)^{2j_1-m_1-m_3}\phi^{j_1}_{m}\phi^{j_1}_{m'}(2j_1+1)\mat{ccc}{j_1&j_1&j_1\\m_1&m&m_3}\mat{ccc}{j_1&j_1&j_1\\-m_1&m'&-m_3} \nonumber\\
  &=\sum_{j_1,m_1}(-1)^{j_1-m_1}\phi^{j_1}_{m_1}\phi^{j_1}_{-m_1}. 
\end{align}
This term is simply the quadratic term of the AR action~\eqref{eq:actionAR}.
Note that this contribution is independent of the solution $T^{m_1m_2m_3}_{j_1j_2j_3}$ and therefore it does not impose any restriction on the homogeneous solution to be considered.

\vspace{10pt}
\paragraph{Terms $TT\psi\psi$\\}

There are four terms of type $TT\psi\psi$. Each of them contributes to the effective action as
\begin{align}
     &\sum_{\substack{m_1,\cdots,m_6\\j_1,\cdots,j_6}}\sixjsim (-1)^{\sum_i(j_i-m_i)}T^{-m_1,-m_2,-m_3}_{j_1j_2j_3}T^{m_3,m_5,-m_4}_{j_3j_5j_4} \nonumber\\
	  &\times\delta^{j_2,0}\delta_{m_2,0}\delta^{j_4,j_6}\psi^{j_4}_{m_4,-m_6}\delta^{j_5,0}\delta_{m_5,0}\delta^{j_6,j_1}\psi^{j_6}_{m_6,m_1} \nonumber\\
	  =& \sum_{\substack{m_1,m_3,m_4,m_6\\j_1,j_3,j_4,j_6}}\matcurl{ccc}{j_1&0&j_3\\j_4&0&j_6}(-1)^{\sum_{i\neq2,5}(j_i-m_i)}T^{-m_1,0,-m_3}_{j_1,0,j_3}T^{m_3,0,-m_4}_{j_3,0,j_4}\delta^{j_4,j_6}\delta^{j_6,j_1}\psi^{j_1}_{m_4,-m_6}\psi^{j_1}_{m_6,m_1} \nonumber\\
	  =&\sum_{j_1,m_1,m_6,m_4}\left[\sum_{m_3}(-1)^{-m_3-m_4}T^{-m_1,0,-m_3}_{j_1,0,j_1}T^{m_3,0,-m_4}_{j_1,0,j_1}\right](-1)^{2j_1-m_1-m_6}\psi^{j_1}_{m_4,-m_6}\psi^{j_1}_{m_1,m_6} 
\end{align}

Thus if the homogeneous solution $T^{m_1m_2m_3}_{j_1j_2j_3}$ is such that
\begin{equation}
    \sum_{m_3}(-1)^{-m_3-m_4}T^{-m_1,0,-m_3}_{j_1,0,j_1}T^{m_3,0,-m_4}_{j_1,0,j_1}=c_{1,j_1} \delta_{m_1,-m_4} \label{eq:TTpsipsicon}
\end{equation}
for some coefficients $c_{1,j_1}$ then we get
\begin{align}
    	  &\sum_{\substack{m_1,\cdots,m_6\\j_1,\cdots,j_6}}\sixjsim (-1)^{\sum_i(j_i-m_i)}T^{-m_1,-m_2,-m_3}_{j_1j_2j_3}T^{m_3,m_5,-m_4}_{j_3j_5j_4}\nonumber \\
	  &\times\delta^{j_2,0}\delta_{m_2,0}\delta^{j_4,j_6}\psi^{j_4}_{m_4,-m_6}\delta^{j_5,0}\delta_{m_5,0}\delta^{j_6,j_1}\psi^{j_6}_{m_6,m_1} \nonumber\\
	  &=\sum_{j_1,m_1,m_6,m_4}\left[\sum_{m_3}(-1)^{-m_3-m_4}T^{-m_1,0,-m_3}_{j_1,0,j_1}T^{m_3,0,-m_4}_{j_1,0,j_1}\right](-1)^{2j_1-m_1-m_6}\psi^{j_1}_{m_4,-m_6}\psi^{j_1}_{m_1,m_6}\nonumber\\
\end{align}

And we specializing to the perturbation~\eqref{eq:psiphirelation} we get
\begin{align}
    &\sum_{\substack{m_1,\cdots,m_6\\j_1,\cdots,j_6}}\sixjsim (-1)^{\sum_i(j_i-m_i)}T^{-m_1,-m_2,-m_3}_{j_1j_2j_3}T^{m_3,m_5,-m_4}_{j_3j_5j_4} \nonumber\\
	  &\times\delta^{j_2,0}\delta_{m_2,0}\delta^{j_4,j_6}\psi^{j_4}_{m_4,-m_6}\delta^{j_5,0}\delta_{m_5,0}\delta^{j_6,j_1}\psi^{j_6}_{m_6,m_1} \nonumber\\
   &=\sum_{j_1,m_1}c_{1,j_1}(-1)^{j_1-m_1}\phi^{j_1}_{m_1}\phi^{j_1}_{-m_1}.
\end{align}
which is the kinetic term of the AR model.

Given a homogeneous solution, the proportionality coefficient $c_{1,j_1}$ can be explicitly computed. Later, we will obtain another condition given by equation~\eqref{eq:tpsipsipsicon} that will be stronger than condition~\eqref{eq:TTpsipsicon} obtained here. Thus, this condition will be automatically satisfied when Equation~\eqref{eq:tpsipsipsicon} is. 

\vspace{10pt}
\paragraph{Term $T\psi T\psi$\\}

The remaining two quadratic contributions from the interaction term of the Boulatov model are of the form $T\psi T\psi$. Each of these terms contributes to the effective action as
\begin{align}
    &\sum_{\substack{m_1,\cdots,m_6\\j_1,\cdots,j_6}}(-1)^{\sum_i(j_i-m_i)}T^{-m_1,-m_2,-m_3}_{j_1j_2j_3}\delta^{j_5,0}\delta_{m_5,0}\delta^{j_3,j_4}\psi^{j_3}_{m_3,-m_4} \nonumber\\
	  & \times T^{m_4,m_2,-m_6}_{j_4j_2j_6}\delta^{j_5,0}\delta_{m_5,0}\delta^{j_1,j_6}\psi^{j_6}_{m_6m_1}\sixjsim \nonumber\\
  &=\sum_{\substack{m_1,m_3,m_4,m_6\\j_1,j_3}}(-1)^{j_1+j_3-m_4-m_6}\psi^{j_3}_{m_3,-m_4}\psi^{j_1}_{m_6,m_1} \frac{1}{\sqrt{(2j_1+1)(2j_3+1)}} \nonumber\\
  &\times \sum_{j_2,m_2}(-1)^{\sum_{i=1}^3(2j_i-m_i)}T^{-m_1,-m_2,-m_3}_{j_1,j_2,j_3}T^{m_4,m_2,-m_6}_{j_3,j_2,j_1}. 
\end{align}
For a general solution of the equation of motion, this term leads to a non-diagonal kinetic term for the $\psi$ field. If the homogeneous solution satisfies the condition
\begin{equation}
      \sum_{j_2,m_2}(-1)^{\sum_{i=1}^3(2j_i-m_i)}T^{-m_1,-m_2,-m_3}_{j_1,j_2,j_3}T^{m_4,m_2,-m_6}_{j_3,j_2,j_1}=c_{2,j_1}c_{2,j_3}\delta_{m_1,-m_6}\delta_{m_3,m_4},
  \label{eq:tpsitpsicon}
\end{equation}
then this contribution becomes
\begin{align}
       &\sum_{\substack{m_1,m_3,m_4,m_6\\j_1,j_3}}(-1)^{j_1+j_3-m_4-m_6}\psi^{j_3}_{m_3,-m_4}\psi^{j_1}_{m_6,m_1} \frac{1}{\sqrt{(2j_1+1)(2j_3+1)}} \nonumber\\
  &\times \sum_{j_2,m_2}(-1)^{\sum_{i=1}^3(2j_i-m_i)}T^{-m_1,-m_2,-m_3}_{j_1,j_2,j_3}T^{m_4,m_2,-m_6}_{j_3,j_2,j_1} \nonumber\\
  =&\left[\sum_{j_1,m_1}(-1)^{j_1-m_1}\frac{c_{2,j_1}}{\sqrt{2j_1+1}}\psi^{j_1}_{m_1,-m_1}\right]^2.
\end{align}

When specializing to the type of perturbation given by equation~\eqref{eq:psiphirelation}, we get
\begin{align}
      \sum_{j_1,m_1}(-1)^{j_1-m_1}\frac{c_{2,j_1}}{\sqrt{2j_1+1}}\psi^{j_1}_{m_1,-m_1}=&\sum_{j_1,m_1,m}(-1)^{j_1-m_1}c_{2,j_1}\phi^{j_1}_{m}\mat{ccc}{j_1&j_1&j_1\\m_1&m&-m_1}, \nonumber\\
  =&\sum_{j_1}c_{2,j_1}\phi^{j_1}_{0}\delta_{j_1,0}\sqrt{2j_1+1}, \nonumber\\
  =&c_{2,0}\phi^0_0.
\end{align}
where we used the equation~\eqref{eq:3jm3eq0}. Therefore, the quadratic term obtained from the $T\psi T\psi$ term can also be made diagonal under the right choice of homogeneous solution and perturbations.

\subsubsection{Cubic terms} 

There is only one type of cubic contribution which comes from the interaction term of the Boulatov model. These terms take the form $T\psi\psi\psi$; there are four such terms and they each contribute as follows:
\begin{align}
     & \sum_{\{j,m\}}(-1)^{\sum\limits_i (j_i-m_i)}T^{-m_1,-m_2,-m_3}_{j_1,j_2,j_3}\delta^{j_5,0}\delta_{m_5,0}\psi^{j_3}_{m_3,-m_4}\delta^{j_2,0}\delta_{m_2,0}\psi^{j_4}_{m_4,-m_6}\delta^{j_5,0}\delta_{m_5,0}\psi^{j_6}_{m_6,m_1}\sixjsim \nonumber\\
 &= \sum_{\substack{m_1,m_3,m_4,m_6\\j_1}}(-1)^{-\sum\limits_{i\neq 2,5}-m_i}T^{-m_1,0,-m_3}_{j_1,0,j_1}\frac{(-1)^{2j_1}}{2j_1+1}\psi^{j_1}_{m_3,-m_4}\psi^{j_1}_{m_4,-m_6}\psi^{j_1}_{m_6,m_1}.
\end{align}
If we impose that the homogeneous solution $T$ satisfies
\begin{equation}
      T^{-m_1,0,-m_3}_{j_1,0,j_1} = c_{3,j_1}(-1)^{-m_3}\delta_{m_1,-m_3}, \label{eq:tpsipsipsicon}
\end{equation}
for some coefficient $c_{3,j_1}$, this contribution becomes
\begin{align}
     &\sum_{\substack{m_1,m_3,m_4,m_6\\j_1}}(-1)^{-\sum_{i\neq 2,5}m_i}T^{-m_1,0,-m_3}_{j_1,0,j_1}\frac{(-1)^{2j_1}}{2j_1+1}\psi^{j_1}_{m_3,-m_4}\psi^{j_1}_{m_4,-m_6}\psi^{j_1}_{m_6,m_1} \nonumber\\
 &=\sum_{\substack{m_3,m_4,m_6\\j_1}}(-1)^{2j_1-m_3-m_4-m_6}\frac{c_{3,j_1}}{2j_1+1}\psi^{j_1}_{m_3,-m_4}\psi^{j_1}_{m_4,-m_6}\psi^{j_1}_{m_6,-m_3}.\\
\end{align}

And when specializing to perturbation~\eqref{eq:psiphirelation}, this contribution becomes
\begin{align}
    &\sum_{\substack{m_1,m_3,m_4,m_6\\j_1}}(-1)^{-\sum_{i\neq 2,5}m_i}T^{-m_1,0,-m_3}_{j_1,0,j_1}\frac{(-1)^{2j_1}}{2j_1+1}\psi^{j_1}_{m_3,-m_4}\psi^{j_1}_{m_4,-m_6}\psi^{j_1}_{m_6,m_1} \\
     &= \sum_{\substack{m_3,m_4,m_6\\j_1}}(-1)^{2j_1-m_3-m_4-m_6}\frac{c_{3,j_1}}{2j_1+1}\sum_{m,m',m''}\phi^{j_1}_{m}\phi^{j_1}_{m'}\phi^{j_1}_{m''} \nonumber\\
 &\times\mat{ccc}{j_1&j_1&j_1\\m_3&m&-m_4}\mat{ccc}{j_1&j_1&j_1\\m_4&m'&-m_6}\mat{ccc}{j_1&j_1&j_1\\m_6&m''&-m_3} \nonumber\\
 &\times (-1)^{j_1}\sum_{m_3,m_4,m_6}(-1)^{3j_1-m_3-m_4-m_6}\mat{ccc}{j_1&j_1&j_1\\m&-m_4&m_3}\mat{ccc}{j_1&j_1&j_1\\m_6&m''&-m_3}\mat{ccc}{j_1&j_1&j_1\\-m_6&m_4&m'} \nonumber\\
 &=\sum_{\substack{m,m',m''\\j_1}}\frac{c_{3,j_1}}{2j_1+1} \matcurl{ccc}{j_1&j_1&j_1\\j_1&j_1&j_1}\phi^{j_1}_{m}\phi^{j_1}_{m'}\phi^{j_1}_{m''}\mat{ccc}{j_1&j_1&j_1\\m&m'&m''}.
\end{align}
Where we have used equations~\eqref{eq:3jpermu} and~\eqref{eq:3jsum6j}, and the fact that $(-1)^{2j_1}=1$ since $j_1$ here has to be an integer for the $3j$ symbol not to vanish. Thus, when imposing the condition~\eqref{eq:tpsipsipsicon}, we get a contribution which corresponds to the interaction term of the AR model. 

Furthermore, as mentioned above, when comparing the two conditions~\eqref{eq:TTpsipsicon} and~\eqref{eq:tpsipsipsicon}, we see that the former will be automatically satisfied when the later is as the two coefficients are related through the relation
\begin{equation}
      c_{1,j_1}=c_{3,j_1}^2. \label{eq:c1c3rel}
\end{equation}

\subsection{Emergence of the Amit-Roginsky-like model}

Now we are ready to extract AR model from the Boulatove action~\eqref{eq:actionTchispin}, based on the two conditions \eqref{eq:tpsitpsicon} and~\eqref{eq:tpsipsipsicon} we discussed in the last subsection. Our main result is the effective action~\eqref{eq:actionphi0epsilon} and~\eqref{eq:actionphijepsilon} for each mode $\phi^j_m$ of the perturbation field (defined through equation \eqref{eq:psiphirelation}). We can see that the form of these actions is the same as the AR one~\cite{Amit:1979ev,Benedetti:2020iku}. 

\paragraph{The effective action for the perturbation $\psi$\\} 

For a perturbation of the form given by equation~\eqref{eq:psiphirelation}, it follows from the previous paragraph that the conditions~\eqref{eq:tpsitpsicon}and~\eqref{eq:tpsipsipsicon} are satisfied. The effective action for the vector perturbation $\phi^{j}_{m}(\vec{\chi})$ then becomes
\begin{equation}
      S[\phi^{j}_{m}]= S_0[\phi^{0}_{0}]+\sum_{j>0}S_j[\phi^{j}_{m}],
      \label{eq:action_phi}
\end{equation}
where 
\begin{equation}
      S_0[\phi^0_0]= \int\dd^3\vec{\chi}\left(\frac{1}{2}\left\{(\nabla \phi^{0}_{0})^2+\left[\mu^2+\frac{\lambda}{3!}(2c_{3,0}^2+c_{2,0}^2)\right](\phi^{0}_{0})^2\right\}-\frac{\xi\lambda}{3!} c_{3,0}\left(\phi^0_0\right)^3 \right),
\end{equation}
and
\begin{align}
      S_j[\phi^{j}_{m}] &=\int\dd^3\vec{\chi}\left\{\frac{1}{2}\left[|\nabla\phi^{j}_{n}|^2+\left(\mu^2+\frac{\lambda}{3!}c_{3,j}^2\right)|\phi^{j}_{n}|^2\right]\right. \nonumber \\
  &\left.-\frac{c_{3,j_1}}{2d_{j}}\frac{\xi\lambda}{3!}\matcurl{ccc}{j&j&j\\j&j&j}\sum_{m_1,m_2,m_3}\phi^{j}_{m_1}\phi^{j}_{m_2}\phi^{j}_{m_3}\mat{ccc}{j&j&j\\m_1&m_2&m_3}\right\}, \label{eq:actionphij}
\end{align}
where $\sum_n|\phi^j_n|^2=\sum_n(-1)^{j-n}\phi^j_n\phi^j_{-n}$. 
The vector fields with different spin labels $j$ decouple and each of them has the form of an AR action with $j$-dependent mass term and coupling. And again, the coefficients $c_{2,j}$ and $c_{3,j}$ can be given explicitly after substituting solutions \eqref{eq:Tfsolspin} and \eqref{eq:PW_eps}.

\vspace{10pt}
\paragraph{Computing coefficients $c_i$ and checking compatibility conditions.\\}

We compute explicitly here the coefficients $c_{1,j}$,$c_{3,j}$ and $c_{2,j}$ for the homogeneous solution~\eqref{eq:Tfsolgroup} to check that these conditions are compatible with our homogeneous solution. Substituting~\eqref{eq:Tfsolgroup} into condition~\eqref{eq:tpsipsipsicon}, we have
\begin{equation}
      \mu\sqrt{\frac{3!}{\lambda}}d_{j_1}f^0_{00}\mat{ccc}{j_1&0&j_1\\-m_1&0&-m_3}=\mu \sqrt{\frac{3!d_{j_1}}{\lambda}}f^0_{00}(-1)^{j_1+m_3}\delta_{m_1,-m_3}=c_{3,j_1}(-1)^{-m_3}\delta_{m_1,-m_3},
\end{equation}
which leads to
\begin{equation}
  c_{3,j}= \begin{cases} (-1)^j \mu \sqrt{\frac{3!d_{j}}{\lambda}}f^0_{00} \hspace{5pt}&\text{if} \hspace{5pt} j\in\mathbb{N} \\ 0 \hspace{5pt}&\text{otherwise}\end{cases}.
\end{equation}

On the other hand, condition~\eqref{eq:tpsitpsicon} yields
\begin{align}
     &\frac{3!\mu^2}{\lambda} \sum_{j_2,m_2}(-1)^{\sum_{i=1}^3(4j_i-m_i)}d_{j_1}d_{j_3}\sum_{n_2,l_2}f^{j_2}_{-m_2,-n_2}f^{j_2}_{m_2,l_2}\mat{ccc}{j_1&j_3&j_2\\m_1&m_3&n_2} \mat{ccc}{j_1&j_3&j_2\\-m_6&m_4&l_2} \nonumber\\
  &=c_{2,j_1}c_{2,j_3}\delta_{m_1,-m_6}\delta_{m_3,m_4},
\end{align}
which leads to the condition for $f^{j_2}_{m_2n_2}$
\begin{equation}
  \sum_{m_2}(-1)^{n_2-m_2}f^{j_2}_{-m_2,-n_2}f^{j_2}_{m_2,l_2}=d_{j_2}c_{f,j_2}^2\delta_{n_2,l_2},
\end{equation}
for some new constants $c_{f,j_2}$. Together with the normalisation condition~\eqref{eq:fnorcon} for $f^j_{mn}$, we get the condition that these new constants should satisfy
\begin{align}
	1&= \sum_{j_2,m_2,n_2,l_2}(-1)^{n_2-m_2}f^{j_2}_{-m_2,-n_2}f^{j_2}_{m_2,l_2} \delta_{n_2,l_2}, \nonumber\\
	&=\sum_{j_2}d^2_{j_2} c_{f,j_2}^2,
\end{align}
and we can get the explicit form \eqref{eq:cfepsilon} of $c_{f,j_2}$ by substituting the heat kernel regularized solution \eqref{eq:PW_eps}.

\vspace{10pt}
\paragraph{The heat kernel regularized solution\\}

The check on the extra conditions performed above on the homogeneous solution~\eqref{eq:PW_eps} still holds at first order in $\epsilon$ when considering the heat kernel regularized solution~\eqref{eq:Tepsilonsolgroupreg}. Using its Peter-Weyl coefficients \eqref{eq:Deltavarepsiloncoef}, we see that
the constant $c_{3,j}$ is simply
\begin{equation}
      c_{3,j}=(-1)^{j} \mu \sqrt{\frac{3!d_j}{\lambda}}(\Delta_\varepsilon)^0_{00}=(-1)^{j}\mu\sqrt{\frac{3!d_j}{\lambda}}\alpha_\varepsilon.
\end{equation}
And the coefficients $c_{f,j}$ would have the form
\begin{equation}
      c_{f,j}=\alpha_\varepsilon  \ee^{-\varepsilon C_j}. \label{eq:cfepsilon}
\end{equation}

Similarly, the condition~\eqref{eq:tpsitpsicon} is only satisfied approximately at first order in $\varepsilon$. Indeed at first order in $\varepsilon$  the Equation~\eqref{eq:3jsumdelta} gives
\begin{equation}
    \sum_{j,m}d_j\ee^{-2\varepsilon C_j}\mat{ccc}{j_1&j_2&j\\m_1&m_2&m}\mat{ccc}{j_1&j_2&j\\m_1'&m_2'&m}\approx\delta_{m_1'm_1}\delta_{m_2'm_2}. \label{eq:3jsumdeltaappro}
\end{equation}
Hence the coefficients $c_{2,j}$ of the condition~\eqref{eq:tpsitpsicon} can then be determined as 
\begin{equation}
 c_{2,j}=\mu d_j\alpha_\varepsilon\sqrt{\frac{3!}{\lambda}}.
\end{equation}
It follows that the effective action for the heat kernel regularized homogeneous solution is
\begin{align}
       S_0[\phi^0_0]&= \int\dd^3\vec{\chi}\left\{\frac{1}{2}\left[(\nabla \phi^{0}_{0})^2+\mu^2\left(1+3\alpha_\varepsilon^2\right)(\phi^{0}_{0})^2\right]-\frac{\sqrt{\lambda}\xi\mu\alpha_\varepsilon}{\sqrt{3!}} \left(\phi^0_0\right)^3 \right\}, \label{eq:actionphi0epsilon}\\
   S_j[\phi^{j}_{m}]  &=\int\dd^3\vec{\chi}\left\{\frac{1}{2}\left[|\nabla\phi^{j}_{n}|^2+\mu^2\left(1+d_j\alpha_\varepsilon^2 \right)|\phi^{j}_{n}|^2\right]\right. \nonumber \\
  &\left.-\frac{(-1)^j}{\sqrt{3!}}\frac{\sqrt{\lambda}\xi\mu\alpha_\varepsilon}{2\sqrt{d_{j}}}\matcurl{ccc}{j&j&j\\j&j&j}\sum_{m_1,m_2,m_3}\phi^{j}_{m_1}\phi^{j}_{m_2}\phi^{j}_{m_3}\mat{ccc}{j&j&j\\m_1&m_2&m_3}\right\}, \label{eq:actionphijepsilon}
\end{align}
where the second equation is exactly the AR action for spin $j$, with mass and interaction coupling dependent on the fundamental GFT coupling and on the spin index $j$. 

This shows that the AR model can be obtained as a particular two dimensional perturbation around classical solutions of the Boulatov model, provided that the classical solution satisfies the conditions given by Equations~\eqref{eq:tpsitpsicon} and~\eqref{eq:tpsipsipsicon}. This is our main result.

Before analysing the resulting generalized AR model further, let us add a few comments on our result. As recalled earlier, the AR model is a vector model on flat euclidean space. From a quantum gravity point of view, the two key ingredients of the model that one would consider challenging to reproduce from the fundamental quantum dynamics are the background flat space it lives on and its local nature. This is because the fundamental formulation of the theory, here the extended Boulatov model with its simplicial quantum gravity underpinning, does not feature continuum spacetime manifold nor local fields defined on it, so both have to be somehow reconstructed in the continuum limit, and the whole framework is diffeomorphism invariant. In our derivation, these issues are apparently bypassed in few simple steps: the continuum limit is encoded in the mean field treatment of the Boulatov model, effectively resumming an infinite series of perturbative, lattice-dependent amplitudes; the desired flat geometry is provided by the homogeneous background solution we expand around; the local characterization of the GFT field perturbations, interpreted as a local vector field in that flat space, is allowed by the extra frame degrees of freedom, in turn coming from scalar matter in the discrete gravity picture, thus a material reference frame. While this ensures some coherence between the interpretation of all the various formal ingredients in our derivation and its result, it is clear that each of them requires further analysis.  

\section{Melonic dominance}\label{melonicdominance}

As already mentioned above, an important feature of the AR model is the dominance of melonic graphs in the large $N=2j+1$ limit.
However, the main difference between the effective action~\eqref{eq:action_phi} and the original AR action
is the presence of the sum over spins $j$. Thus we have to check whether or not this new summation spoils the existence of a melonic regime.  Even though the general behaviour of $\{3nj\}$ symbols as functions of $j$ is an open issue \cite{Haggard:2010,Costantino,Bonzom:2012,Don:2018}, one can qualitatively study the behaviour of the Feynman amplitudes of the model and give additional constraints to ensure the existence of such melonic regime.

\subsection{Feynman amplitudes for the non-regularized solution}

For simplicity, we will drop  below  the heat kernel regularisation and work with the actions given by Equation~\eqref{eq:actionphij}, including the sum over spin labels $j$. As in the AR model, each Feynman diagram $\gamma$ of our new model consists of isoscalar part $I_{\gamma}$ and isospin part $A_{\gamma}$~\cite{Amit:1979ev,Benedetti:2020iku}:
\begin{equation}
    \mathcal{A}_{\gamma}=\sum_{j}c_{\gamma}\left(\frac{\lambda \{6j\}}{3!(2j+1)}\right)^{v}I_{\gamma}A_{\gamma},
\end{equation}
where $c_{\gamma}$ is the combinatorial factor of the diagram. The isoscalar part yields a space integral, so one needs to study the isospin part to find how the Feynman amplitude depends on $N$.\\

The melonic graphs are Fully $2$-Particle Reducible (F2PR) diagrams, i.e. they always admit a $2$-cut which gives another melonic graph with fewer vertices, until the trivial graph is reached. Their contribution writes
\begin{equation}
    \mathcal{A}_{F2PR}\sim\sum_{j}(2j+1)^{1-3n}\{6j\}^{2n}\equiv\bar{\mathcal{A}}_{F2PR},
\end{equation}
with for a graph with $v=2n$ vertices. For a graphs which is not F2PR, the Feynman amplitude can be factorized as a product of $2$-particule irreducible graphs
\begin{equation}
    \mathcal{A}_{NF2PR}\sim\sum_{j}(2j+1)^{-n_0-2n}\prod_{i=1}^k A_{\{3n_i j\}}\{6j\}^{2n}
    \equiv\bar{\mathcal{A}}_{NF2PR},
\end{equation}
where
\begin{equation}
    n=1+n_0-k+\sum_{i=1}^k n_i,
\end{equation}
and $A_{\{3n_i j\}}$ is the amplitude of a three-particle irreducible diagrams with $2n_i$ vertices.

When $N=2j+1$ goes to infinity, the amplitudes $\bar{\mathcal{A}}_{NF2PR}$ is conjectured to obey the following bound~\cite{Amit:1979ev}
\begin{equation}\label{ln}
    \bar{\mathcal{A}}_{NF2PR}\leq \sum_j (2j+1)^{1-3n-\alpha}\{6j\}^{2n},
\end{equation}
for some real number $\alpha>0$. Asymptotically, when $N\rightarrow \infty$, both $n\geq 1$ and the $6j$ symbol are small with respect to $N$. Therefore we get the following bound
\begin{equation}
    \bar{\mathcal{A}}_{F2PR}<\sum_jN^{1-3n}=\sum_j (2j+1)^{1-3n}=\left(1-2^{1-3n}\right)\zeta(3n-1),
\end{equation}
where $\zeta$ is the Riemann zeta function, which is a monotonically decreasing finite function of $n$. 

If one assumes that the bound~\eqref{ln} holds for any value of $N$, then the amplitude of a $NF2PR$ graphs is also finite. If the bound~\eqref{ln} fails to hold for values of $N$ satisfying $N<N_t$ for some bound $N_t$, then the sum from $N=3$ ($j$ can only be an integer no smaller than $1$, so $N\geq 3$) to $N=N_t$ is still a finite number, while the sum from $N=N_t$ is finite as well. Therefore, it is possible that $\mathcal{A}_{NF2PR}$ is comparable with $\mathcal{A}_{F2PR}$ since the maximal value of $\zeta(3n-1)$ is only $\pi^2/6\simeq 1.645$. 

One can thus conclude that the sum over $j$ can dramatically change the amplitude of a Feynman graphs of the AR model and spoil the melonic limit at large $N$. However, one can find ways to rule out this possibility and ensure that the melonic dominance is preserved. This will be illustrated in the following subsection.

\subsection{Restoring the melonic dominance}

One na\"ive way to  restore the melonic dominance is of course to further specialize the form of the perturbation~\eqref{eq:Tpsispin} in order to enforce the selection of one single value for the spin $j$, thus getting rid of the sum over spin labels and leading to the original AR model:
\begin{equation}
      (T_\psi)^{m_1m_2m_3}_{j_1j_2j_3}(\vec{\chi})=T^{m_1m_2m_3}_{j_1j_2j_3}+\delta^{j_1j}\delta^{j_2,0}\delta_{m_2,0}\psi^{j_1}_{m_1,m_3}(\vec{\chi}), 
\end{equation}

Another, more interesting, way to recover melonic dominance is to work with the approximate solution~\eqref{eq:Tepsilonsolgroupreg}. Indeed, when $j_2=0$ the solution has the form:
\begin{equation}
      (T_{\varepsilon})^{m_1m_2m_3}_{j_1j_2j_3}=\mu\alpha_{\varepsilon}\sqrt{\frac{3!}{\lambda}}\ee^{-2\varepsilon C_{j_1}} \sqrt{2j_1+1}(-1)^{j_1-m_1}\delta_{j_1,j_3}\delta_{m_1,-m_3}.
\end{equation}
For $\varepsilon=(2j_{\mathrm{max}}(j_{\mathrm{max}}+1))^{-1}$, the expression
above scales as $\sqrt{2j_1+1}$ for $j_1<j_{\mathrm{max}}$, with $j_{\mathrm{max}}$ a large number. Hence, in the Peter-Weyl expansion, the coefficients with larger $j$ are dominant, and the coefficients $(T_{\varepsilon})^{m_1m_2m_3}_{j_1j_2j_3}$ with $j_i< j_{\mathrm{min}}$ for some threshold $j_{\mathrm{min}}$ can be neglected. We require that $j_{\mathrm{min}}$ is also a large number so that the bound~\eqref{ln} is valid.
At first order in $\epsilon$ one then has:
\begin{equation}
      (T_\psi)^{m_1m_2m_3}_{j_1j_2j_3}(\vec{\chi})\simeq
      \left\{\begin{array}{cc}    T^{m_1m_2m_3}_{j_1j_2j_3}+\delta^{j_1j}\delta^{j_2,0}\delta_{m_2,0}\psi^{j_1}_{m_1,m_3}(\vec{\chi}), & j_{\mathrm{min}}\leq j_i\leq j_{\mathrm{max}}\nonumber\\
 0, &\text{otherwise}\end{array}  \right. .
\end{equation}
Such perturbations $\phi^j_m$ will lead to the amplitude
\begin{equation}
      \mathcal{A}_{\gamma}=\sum_{j=j_{\mathrm{min}}}^{j=j_{\mathrm{max}}}c_{\gamma}\left(\frac{\lambda \{6j\}}{3!\sqrt{2j+1}}\right)^{v}I_{\gamma}A_{\gamma},
\end{equation}
which becomes an infinitesimal again for large $j_{\mathrm{min}}$ and $j_{\mathrm{max}}$, while the non-F2PR graphs are higher order infinitesimals as in the original AR model. The melonic dominance is thus restored.

\section{Concluding remarks} \label{sec:summary}

In this work, we have obtained a generalised version of the Amit-Roginski model as a (two-dimensional) perturbation around a classical homogeneous solution of the Boulatov group field theory model for 3d quantum gravity, extended to include (what plays the role, in the discrete gravity picture, of) scalar matter degrees of freedom, which end up providing a material frame and embedding coordinates in the resulting AR model. 
This is an interesting result from a physical point of view, first of all, since it connects 3d quantum gravity, in a well-studied and mathematically rich formulation, and the AR model, itself of great mathematical interest.
The main difference between our effective action for the perturbation and the usual AR model is the presence of the summation on the spin index $j$. While it is still unclear whether this summation could spoil the dominance of melonic diagram in the most general framework, it is possible to preserve this  melonic limit also in this generalised AR model by making use of the heat kernel regularization and taking a double scaling limit. 

It is also an interesting result from a more conceptual point of view, since it shows an example of the emergence of a local field theory (in flat space) from a background independent quantum gravity formalism based on non-spatiotemporal structures (meaning, not corresponding directly to quantized continuum spacetime-based fields), which is an outstanding challenge for most quantum gravity approaches. In fact, our result resonates (especially in the key role played by scalar matter used as a relational frame) with recent work on GFT cosmology \cite{Marchetti:2021gcv}.

\medskip

This result opens the way for at least three different generalisations. Firstly, a natural follow-up would be to find other classical solutions to the Boulatov model and to study perturbations around these solutions to see if they also admit Amit-Roginski-like perturbations. 

Secondly, as already mentioned in the Introduction, the holographic SYK model is another type of field theory that is known to enjoy a melonic limit. 
It thus appears interesting to us to investigate how the SYK model as well can be obtained within a GFT setup, again in terms of fluctuations over non-perturbative quantum gravity configurations.

Third, as we have already mentioned, a detailed comparison of our 3d quantum gravity setting with the $O(N)^3$-symmetric random tensor model studied in \cite{Benedetti:2019sop}, leading to a similar classical solution and another possible route to derive the AR model, would be very interesting to perform.

Once more, while exploring generalizations of our results, many elements in our derivation deserve a deeper and more extensive analysis. Among these, we mention again: the quantum geometric interpretation and effects of the sum over topologies in the GFT construction; the continuum physical interpretation of the classical solutions of GFT equations of motion; the renormalization group flow and continuum limit of the extended TGFT models, with both local and tensorial directions, that were the starting point of our analysis.  

\acknowledgements

The authors acknowledge financial support from the Bordeaux-LMU collaboration grant.
DO acknowledges financial support from the Deutsche Forschung Gemein-schaft (DFG). XP is supported by China Scholarship Council. 
VN and AT have been partially supported by the ANR-
20-CE48-0018 “3DMaps” grant. 
AT is partially supported by the PN 23 21 01 01/2023 grant.
 DO, VN and AT acknowledge support of the Institut Henri Poincaré (UAR 839 CNRS-Sorbonne Université), and LabEx CARMIN (ANR-10-LABX-59-01)
during the 2023 IHP trimester "Quantum gravity, random geometry and holography".

\appendix

\section{Definitions and identities from $SU(2)$ recoupling theory} \label{sec:su2recoupling}

We give several definitions and properties related to $SU(2)$ recoupling theory used in the article. All those properties are classical results on recoupling theory of $SU(2)$, and we refer the interested reader to Ilkka M\"akinen's introduction~\cite{Makinen:2019rou} as well as Pierre Martin-Dussaud's lively note \cite{Martin-Dussaud:2019ypf} on this topic for more details. 

\subsection{Haar measure and Wigner matrices}

From the Peter-Weyl theorem, the Wigner matrices $D^j_{mn}(g)$ form an orthogonal basis of the functions $f:SU(2) \rightarrow \mathbb{C}$. This orthogonality relation is encoded in the Haar measure via the relation
\begin{equation}
      \int\dd g D^j_{mn}(g)\bar{D}^{j'}_{m'n'}(g) =\frac{1}{(2j+1)}\delta^{jj'}\delta_{mm'}\delta_{nn'},
\end{equation}
where the Wigner matrices satisfy
\begin{equation}
      D^j_{mn}(g)=(-1)^{m-n}\bar{D}^j_{-m,-n}(g).
\end{equation}

\subsection{$3j$-symbol and its properties}

The $3j$ symbol is invariant under the action of $SU(2)$ group, 
\begin{equation}
D^{j_1}_{m_1n_1}D^{j_2}_{m_2n_2}D^{j_3}_{m_3n_3}\mat{ccc}{j_1&j_2&j_3\\n_1&n_2&n_3}=\mat{ccc}{j_1&j_2&j_3\\m_1&m_2&m_3}.
\end{equation}

It's also invariant under the even permutations of indices, while it acquires an additional phase under odd permutations
\begin{equation}
       \mat{ccc}{j_1&j_2&j_3\\ m_1&m_2&m_3}=(-1)^{j_1+j_2+j_3}\mat{ccc}{j_2&j_1&j_3\\ m_2&m_1&m_2}. \label{eq:3jpermu}
\end{equation}

The same phase also appear if we replace $m_i$ by their negative
\begin{equation}
       \mat{ccc}{j_1&j_2&j_3\\ -m_1&-m_2&-m_3}=(-1)^{j_1+j_2+j_3}\mat{ccc}{j_1&j_2&j_3\\ m_1&m_2&m_3}. 
\end{equation}

The $3j$ symbols satisfy two orthonormal relations
\begin{align}
     &(2j_3+1)\sum_{m_1,m_2}\mat{ccc}{j_1&j_2&j_3\\m_1&m_2&m_3}\mat{ccc}{j_1&j_2&j_3'\\m_1&m_2&m_3'}=\delta_{j_3,j_3'}\delta_{m_3,m_3'}, \\
  &\sum_{j_3,m_3}(2j_3+1)\mat{ccc}{j_1&j_2&j_3\\m_1&m_2&m_3}\mat{ccc}{j_1&j_2&j_3\\m_1'&m_2'&m_3}=\delta_{m_1,m_2'}\delta_{m_2,m_2'},  \label{eq:3jsumdelta}
\end{align}

Finally, when one of the magnetic moment (say $m_3$) vanishes, then the $3j$ symbol vanishes unless $m_1 = -m_2$ and we have
\begin{equation}
       \sum_m (-1)^{j-m}\mat{ccc}{j&j&k\\m&-m&0}=\sqrt{2j+1}\delta_{k,0}. \label{eq:3jm3eq0}
\end{equation}

And in particular for $k=0$ we have
\begin{equation}
      \mat{ccc}{j_1&0&j_3\\n_1&0&n_3}=\delta^{j_1,j_3}\frac{1}{\sqrt{2j_1+1}}(-1)^{j_1+n_1}\delta_{n_1,-n_3} \label{eq:3jj2eq0} \\
\end{equation}

\subsection{$6j$-symbol and its properties}

The $6j$ symbol is defined as
\begin{align}
      \left\{\begin{array}{ccc}j_1&j_2&j_3\\j_4&j_5&j_6\end{array}\right\}
  =&\sum_{j_i,m_i} (-1)^{\sum_{a=1}^6(j_a-m_a)} \mat{ccc}{j_1&j_2&j_3\\-m_1&-m_2&-m_3}\mat{ccc}{j_1&j_5&j_6\\m_1&-m_5&m_6} \nonumber \\
  &\cdot \mat{ccc}{j_4&j_2&j_6\\m_4&m_2&-m_6}\mat{ccc}{j_4&j_5&j_3\\-m_4&m_5&m_3}.   \label{eq:6jdef}
\end{align}
It enjoys several symmetries properties that we do not make use of in the main body. We refer the interested reader to~\cite{Makinen:2019rou} where they are explicitly mentioned.

Using the $6j$ symbol we have
\begin{align}
       &\sum_{n_1,n_2,n_3}(-1)^{\sum_{a=1}^{3}(k_a-n_a)} \mat{ccc}{j_1&k_2&k_3\\m_1&-n_2&n_3}\mat{ccc}{k_1&j_2&k_3\\n_1&m_2&-n_3}\mat{ccc}{k_1&k_2&j_3\\-n_1&n_2&m_3} \nonumber \\
   &=\left\{\begin{array}{ccc}j_1&j_2&j_3\\k_1&k_2&k_3\end{array}\right\} \mat{ccc}{j_1&j_2&j_3\\ m_1&m_2&m_3}. \label{eq:3jsum6j} 
\end{align}

Finally when one of the spin index (say $j_6$) vanishes we have
\begin{equation}
      \left\{\begin{array}{ccc}j_1&j_2&j_3\\j_4&j_5&0\end{array}\right\}=\frac{\delta_{j_1,j_5}\delta_{j_2,j_4}}{\sqrt{d_{j_1}d_{j_2}}}(-1)^{j_1+j_2+j_3}\{j_1~j_2~j_3\}. \label{eq:6jj6eq0}
\end{equation}


\end{document}